\documentclass[twocolumn, tighten, dvipsnames, floatfix]{aastex63}
\usepackage[]{xcolor}
\usepackage{booktabs}
\usepackage{physics}
\usepackage{longtable}
\usepackage{threeparttablex} % for "ThreePartTable" environment
\usepackage{array}
\usepackage{amsmath}
\usepackage{comment}
\usepackage{changepage}
\usepackage{rotating}
\usepackage{lineno}
\usepackage[flushmargin]{footmisc}
\usepackage{graphicx}
\usepackage{afterpage}
\usepackage{subfigure}
\usepackage{soul} % para riscar o texto

\def\arcsec{\hbox{$^{\hbox{\rlap{\hbox{\lower4pt\hbox{$\,\prime\prime$}}
  }}}$} \ }

\def\arcmin{\hbox{$^{\hbox{\rlap{\hbox{\lower4pt\hbox{$\;\prime$}}
  }\hbox{$\frown$}}}$}}

\usepackage{lipsum}
\usepackage{float}

\shorttitle{Color--Magnitude Relation for BHB Stars}
\shortauthors{Barbosa et al.}

\begin{document}

\title{The SDSS-Gaia View of the Color--Magnitude Relation for Blue Horizontal-branch Stars}

\correspondingauthor{Fabrícia Oliveira Barbosa}
\email{fabriciaob@usp.br}

\author[0000-0002-8262-2246]{Fabr\'icia O. Barbosa}
\affil{Universidade de S\~ao Paulo, Instituto de Astronomia, Geof\'isica e Ci\^encias Atmosf\'ericas, Departamento de Astronomia, \\ SP 05508-090, S\~ao Paulo, Brazil}

\author[0000-0002-7529-1442]{Rafael M. Santucci}
\affiliation{Universidade Federal de Goi\'as, Instituto de Estudos Socioambientais, Planet\'ario, Goi\^ania, GO 74055-140, Brazil}
\affiliation{Universidade Federal de Goi\'as, Campus Samambaia, Instituto de F\'isica, Goi\^ania, GO 74001-970, Brazil}

\author[0000-0001-7479-5756]{Silvia Rossi}
\affil{Universidade de S\~ao Paulo, Instituto de Astronomia, Geof\'isica e Ci\^encias Atmosf\'ericas, Departamento de Astronomia, \\ SP 05508-090, S\~ao Paulo, Brazil}

\author[0000-0002-9269-8287]{Guilherme Limberg}
\affil{Universidade de S\~ao Paulo, Instituto de Astronomia, Geof\'isica e Ci\^encias Atmosf\'ericas, Departamento de Astronomia, \\ SP 05508-090, S\~ao Paulo, Brazil}

\author[0000-0002-5974-3998]{Angeles P\'erez-Villegas}
\affil{Instituto de Astronom\'ia, Universidad Nacional Aut\'onoma de M\'exico, Apartado Postal 106, C. P. 22800, Ensenada, B. C., Mexico}

\author[0000-0002-0537-4146]{H\'elio D. Perottoni}
\affil{Universidade de S\~ao Paulo, Instituto de Astronomia, Geof\'isica e Ci\^encias Atmosf\'ericas, Departamento de Astronomia, \\ SP 05508-090, S\~ao Paulo, Brazil}
\affiliation{Institut de Ciències del Cosmos (ICCUB), Universitat de Barcelona (IEEC-UB), Martí i Franquès 1, E08028 Barcelona, Spain}

\begin{abstract}

We present an updated sample of blue horizontal-branch (BHB) stars selected from the photometric and spectroscopic data from Sloan Digital Sky Survey and its associated project Sloan Extension for Galactic Understanding and Exploration (SEGUE). With this data, we selected candidates for A-type stars in the color-color space and then a mixture modeling technique was implemented in order to distinguish between BHB and main-sequence/blue-straggler stars based on their surface gravity values ($\log \rm{g}$) estimated by the SEGUE Stellar Parameter Pipeline. Our robust approach allows us to attribute individual probabilities of each star truly being in the BHB stage. Hence, our method is advantageous in comparison to previous SEGUE BHB selections that adopted simple $\log \rm{g}$ cuts. 
We also revisit the color--magnitude relation for these stars and propose two calibrations, based on updated distances for Galactic globular clusters, to estimate absolute magnitudes with $(g-r)_0$ and $(u-r)_0$ colors.

\end{abstract}

\keywords{Galaxy: stellar halo -- stars: horizontal branch -- stars: distances}

%\linenumbers

\defcitealias{deason}{D11}

%===============================================
%===============================================
\section{Introduction}
\label{sec:intro}

The Gaia mission \citep{gaia_mission} has provided a better understanding of the Galaxy, in particular regarding the field of Galactic Archaeology \citep{helmi_review, gaia_review}. The astrometric information provided for an unprecedented number of objects has dramatically changed the way we study the Galactic halo \citep[e.g.,][]{belokurov_ges, myeong_seq, koppelman, malhan}.

Despite the huge amount of direct measurements supplied by Gaia, distances inferred from brightness are still of great value. At magnitude $G < 15$, the early third data release (EDR3) presents parallax uncertainties of ${\sim}0.02$ mas \citep{gaia_edr3}, and they increase significantly for fainter stars.
To overcome this limitation, we can use various well-known distance tracers such as RR Lyrae \citep{rrl_is_cepheid}, Cepheids \citep{henrietta}, and blue horizontal-branch (BHB) stars \citep{bhb_as_candle}.

BHBs are metal-poor ([Fe/H]\footnote{\noindent[A/B] $= \log \rm (N_A/N_B)_\star - \log (N_A/N_B)_\odot$, where $\rm N_A$ and $\rm N_B$ are the number density of atoms of the elements A and B, respectively. $\star$ refers to the considered star, and $\odot$ refers to the Sun.} $\lesssim -0.5$; \citealt{santucci_map}) A or B-type stars that burn helium in their cores.
These evolved stars present a high and nearly constant luminosity, making them perfect for investigating the outer regions of the halo and the assembly history of our Galaxy \citep{xue08, xue11, deason, deason17, Belokurov2014_SgrStream, santucci_map}. In recent works, BHBs were used to study dynamical substructures and stellar streams \citep{Yuan2019cetus, yuan_struc_20, Yuan2022cetus, Penarrubia2021sgr, li_streams, sausage_wu}, the connection between the apocenter pile-up of orbits and the so-called ``break-radius'' of the stellar halo density profile \citep{dea_break}, the anisotropy of the halo velocity distribution \citep{Lancaster2019bhbs}, the age gradient of the halo out to $\sim$35 kpc \citep{Whitten2019bhbs}, to estimate the total dynamical mass of the Milky Way \citep{mass_deason, mass_bird}, and even to demonstrate the influence of the Large Magellanic Cloud in our Galaxy's halo \citep{erkal_lmc, petersen_lmc}.

The well-defined structure of the horizontal branch in the color-magnitude diagram (CMD), a roughly constant luminosity, permits the development of a distance calibration for these BHBs. The first approximation developed was a linear fit, using the ($B-V$) color and absolute magnitude in the $V$ band, for stars in globular clusters \citep{hayes}. Likewise, \citet{preston} defined a smoother relation, a fourth degree polynomial, for the same color-magnitude space. Two decades later, a widely used calibration was presented by \citet[][hereafter \citetalias{deason}]{deason} based on magnitudes in the $ugriz$ system for the Sloan Digital Sky Survey \citep[SDSS;][]{sdss} eighth data release \citep[DR8;][]{Aihara2011sdssDR8}, which had its color range extended by \cite{belokurov_bhb} afterwards. 
In the meantime, \cite{fermani} argued that it is extremely important to take into account the effect of the metallicity on the absolute magnitude estimation, proposing a new calibration based on a statistical method. However, \cite{santucci_frac} and \cite{utkin} showed that the differences between considering or not the metallicity in the relations are negligible, with \citetalias{deason}'s estimates being $2.5\%$ higher, within (1-$\sigma$) errors of both calibrations.

\citetalias{deason}'s relation still remains the most used calibration for BHB stars \citep{santucci_frac,santucci_map,exem_dea_18, Whitten2019bhbs, exem_dea_20, exem_dea_22} even though photometric data have been updated several times since then. Moreover, we can now compare photometric distances of BHB stars with purely geometric estimates from Gaia's parallaxes (e.g., \citealt{bailer}) as well as new measurements for Galactic globular clusters \citep{vasi}. These facts bring to light the relevance of reviewing the \citetalias{deason}'s calibration with recent data.

This paper is organized as follows. In Section \ref{sec:data}, we describe the photometric selection and revise a previous method to identify BHB stars. Section \ref{sec:calib} presents the selection of stars in globular clusters and the method used to define the absolute magnitude calibration. Finally, in Section \ref{sec:conc} we discuss our results.

%===============================================
%===============================================
\section{Data}
\label{sec:data}

\subsection{A-type stars}
\label{A_stars}

The initial selection of A-type stars was made using the photometry from the sixteenth data release (DR16) of SDSS \citep{sdssdr16}. For the selection of BHB stars, we were specially interested in the spectroscopic data obtained by the Sloan Extension for Galactic Understanding and Exploration \citep[SEGUE;][]{yanny} processed by the SEGUE Stellar Parameter Pipeline \citep[SSPP;][]{Lee08a,Lee08b}\footnote{\noindent Last run on DR9 \citep{dr9, segue2}.}.

We implemented color cuts applying the following criteria: $-0.3 < (g-r)_0 < 0.1$ and $0.8 < (u-g)_0 < 1.4$, similar to those used in previous works (\citealt{sirko}, \citetalias{deason}). All the magnitudes were corrected using the extinction coefficients ($A_g$, $A_r$, $A_u$) provided by the SDSS catalog itself and we removed stars with relative errors in the $g$-band magnitude greater than 1\%.

The photometric selection is able to exclude several undesired objects, such as white dwarfs, quasars and cooler spectral types \citep{yanny, vickers}, but the major source of contamination, blue straggler stars (BSSs), remains. The distinction between evolved and main-sequence stars/BSSs is commonly made by investigating spectral features, specially Balmer lines, whose depths are affected by effective temperature ($T_{\rm eff}$) and widths by surface gravity ($\log \rm{g}$). With the output of SSPP, we can directly inspect these stellar atmospheric parameters. Therefore, we cross-matched the filtered sample with the SSPP catalog using $5''$ radius. In addition to color filters, we restricted our sample to stars with moderate signal-to-noise ratio ($S/N > 10$) and $7500\,{\rm K} < T_{\rm eff} < 10000\,{\rm K}$ \citep{deason2012, santucci_frac}, where $T_{\rm eff}$ is the estimate adopted by the pipeline. Duplicated stars with the smallest $S/N$ were removed, which resulted in 16463 stars.
The restrictions above remove poor-quality data and cooler stars that could remain after the color cut, which assures that contamination from other non-BSS stars is minimal.

%===============================================
%===============================================
\subsection{BHB stars}
\label{sec:gmm}

One of the techniques used to disentangle BSSs and BHB stars is the $f_m$ versus $D_{0.2}$ method \citep{pier}, where $f_m$ is the minimum flux relative to the continuum level and $D_{0.2}$ is a measurement of the line width of Balmer lines, so it provides indirect information regarding both $T_{\rm eff}$ and $\log \rm{g}$. Later, a different approach was proposed by \cite{method_bc} based on the parameters of the Sérsic profile \citep{sersic}, which describes the shape of the lines.

\begin{figure*}[pt!]
  \begin{minipage}[b]{0.5\textwidth}
    \centering
    \includegraphics[width=\columnwidth, trim={0 2.2cm 0 0},clip]{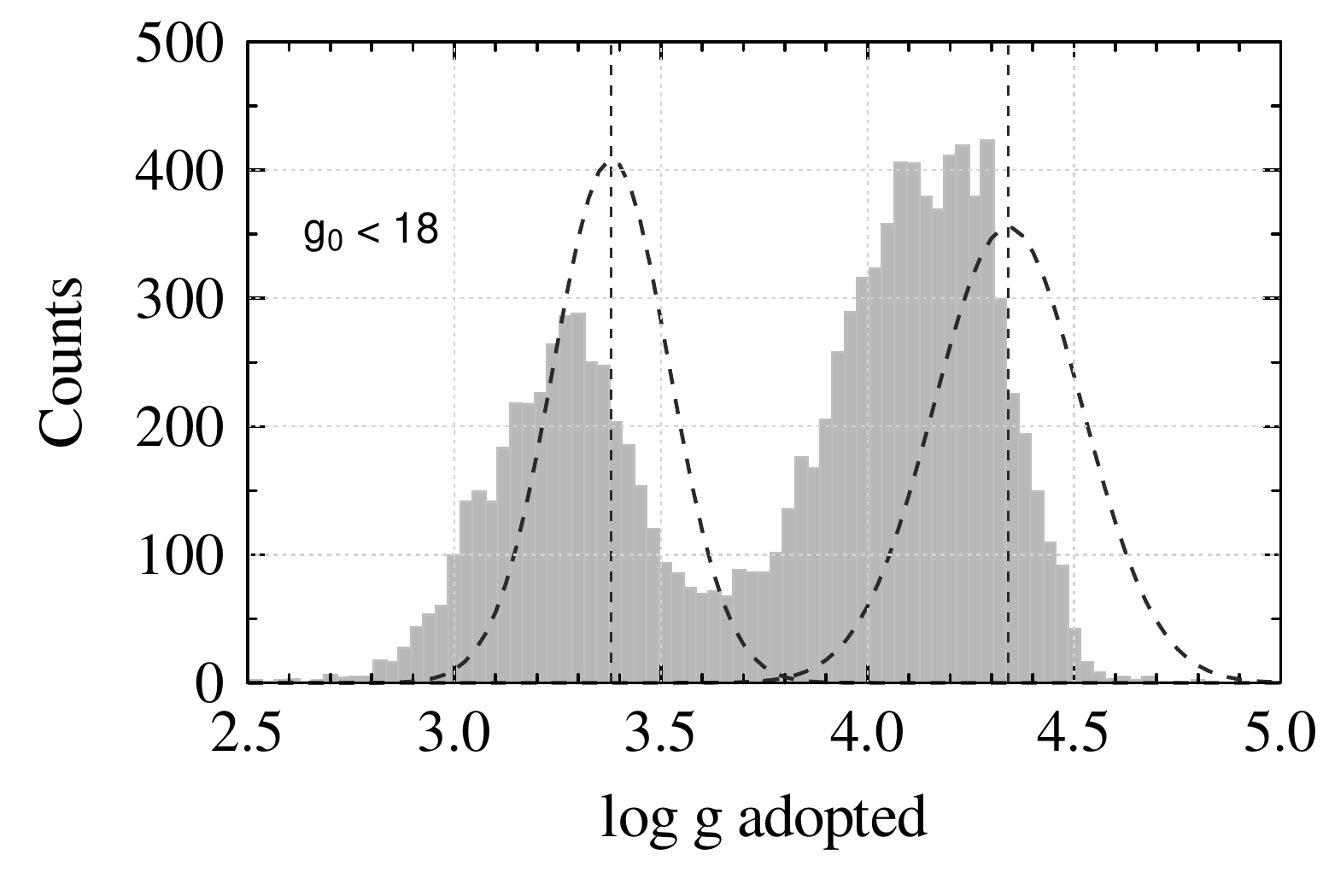}
    \includegraphics[width=\columnwidth, trim={0 0 0 0.4cm},clip]{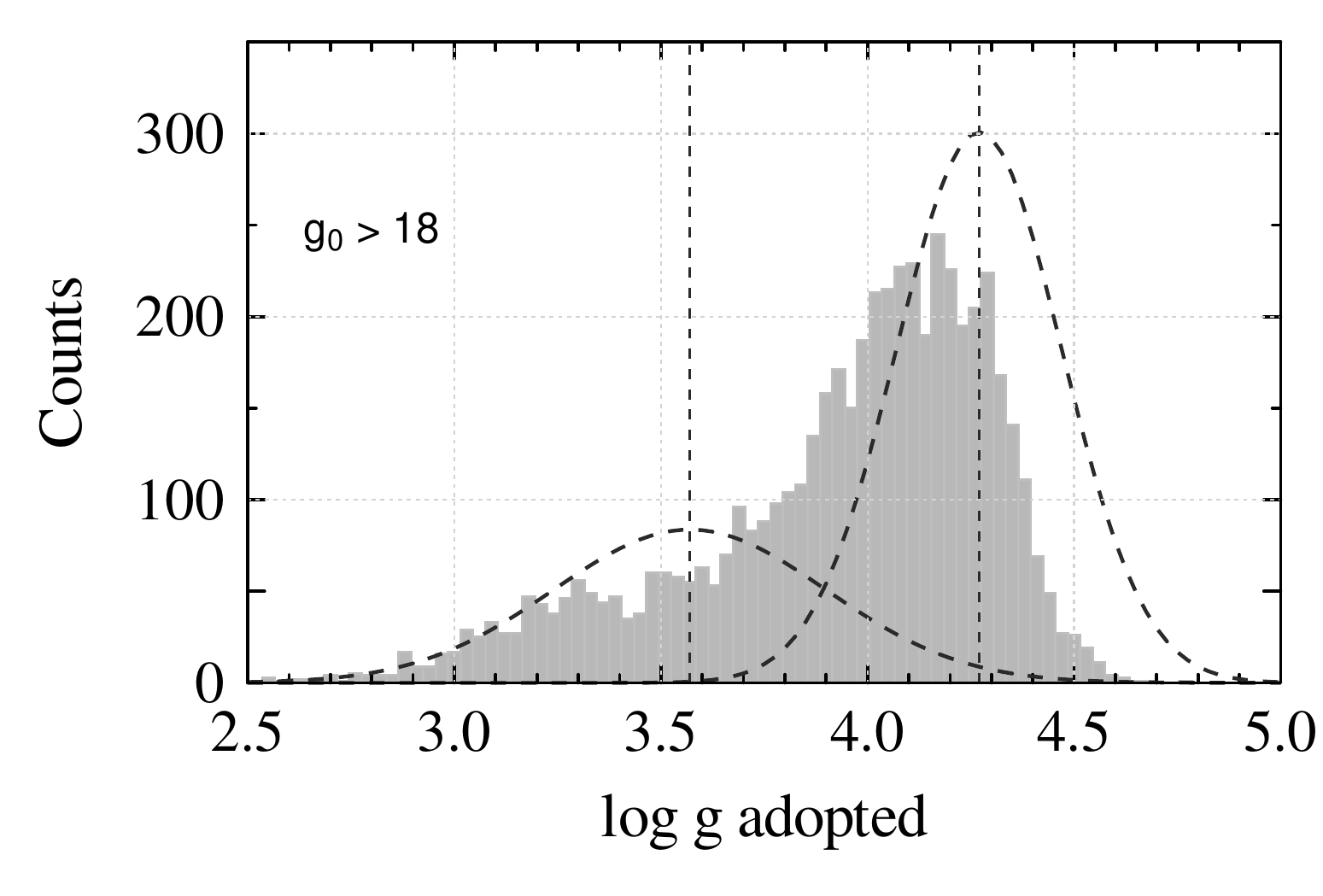}
  \end{minipage}
  \begin{minipage}[b]{0.5\textwidth}
    \centering
    \includegraphics[width=\columnwidth, trim={0 2.2cm 0 0},clip]{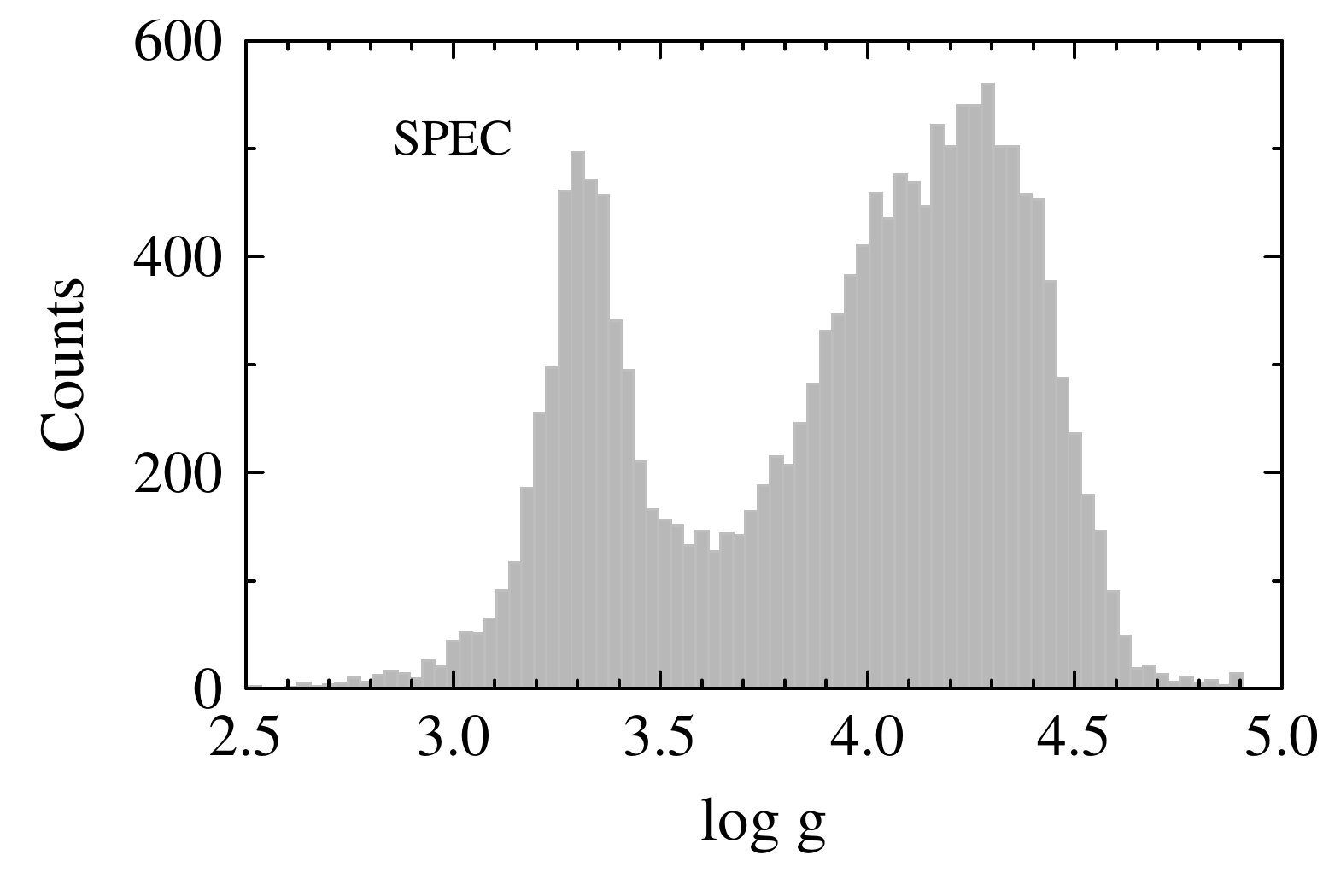}
    \includegraphics[width=\columnwidth, trim={0 0 0 0.4cm},clip]{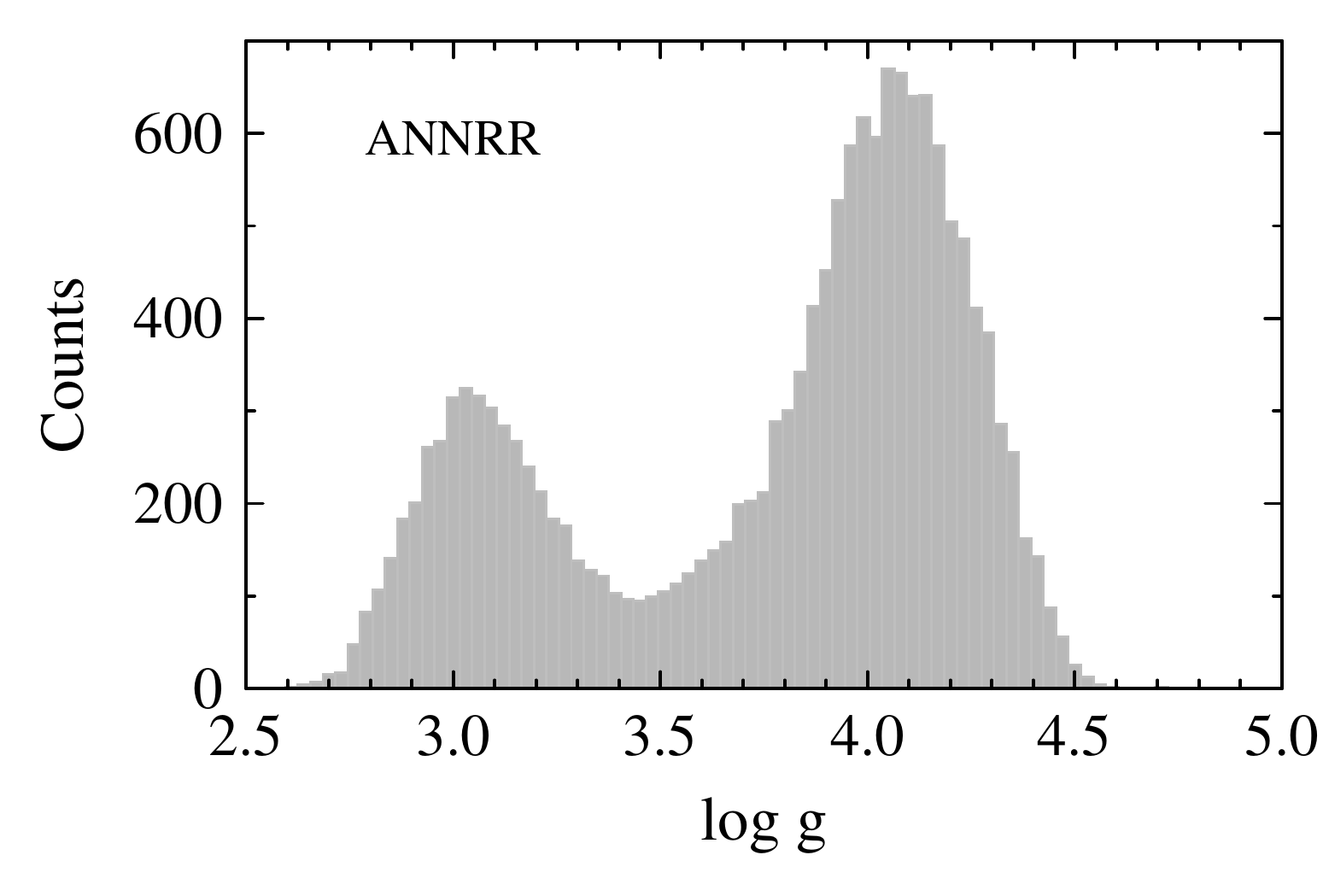}
  \end{minipage}%
  \caption{Histograms of $\log \rm{g}$. Left column: log g adopted by the pipeline for stars with $g_0 < 18$ (top) and $g_0 > 18$ (bottom). Dashed lines are the Gaussian distributions defined by \cite{santucci_frac}. Right column: log g estimates provided by SSPP spectroscopically determined (top) and from \texttt{ANNRR} method (bottom) for all stars.}
  \label{fig:gauss}
\end{figure*}

BSSs present a stronger $\log \rm{g}$ than those located in the horizontal branch. \cite{santucci_frac} showed that these stellar types are clearly distinguishable for magnitudes $g_0 < 18$ with SEGUE/SDSS DR8 data, being possible to classify them by fitting a combination of two Gaussian functions to their $\log \rm{g}$ distributions. This method was proved to be in good concordance with spectral analysis, with more than 90\% of agreement. When replicating this procedure with current SEGUE data, we noticed a change in the peaks of both groups and a greater overlap in the $\log \rm{g}$ distribution as presented in the left column of Fig. \ref{fig:gauss}. This is observed even for relatively bright A-type stars ($g_0 < 18$, top panel), which makes it more difficult to separate these objects with that simple approach (dashed lines indicate the Gaussian fits from \citealt{santucci_frac}). The differences are probably due to changes between the releases of SDSS on the $\log \rm{g}$ estimates considered to obtain the final adopted parameter\footnote{\noindent See \url{https://www.sdss.org/dr16/spectro/sspp_changes/} for detailed information.}.

\subsection{Classification} \label{subsec:class}

Given the two-Gaussian-like morphology of the $\log \rm{g}$ distributions observed for our sample of A-type stars (Fig. \ref{fig:gauss}), we used a Gaussian Mixture Model (GMM) unsupervised approach in order to distinguish BHBs and BSSs. For this task, we utilize the \texttt{scikit-learn} \citep{scikit-learn} \texttt{GaussianMixture}\footnote{\noindent\url{https://scikit-learn.org/stable/modules/generated/sklearn.mixture.GaussianMixture.html\#sklearn.mixture.GaussianMixture}.} package. In this GMM implementation, the expectation-maximization algorithm \citep{EMalgo1977} is employed in the search for the best-fit model. 

The GMM technique fits the data as a finite combination of $K$ Gaussian distributions. As made previously by \cite{santucci_frac}, $K$ was defined based on visual inspection of $\log \rm{g}$ estimates presented in Fig. \ref{fig:gauss} and the assumption that the contamination is predominantly of main-sequence stars/BSSs. Therefore, $K=2$ is an adequate value for the sample.
Moreover, GMM can be readily applied to data of arbitrary dimensionality. Therefore, we take advantage of such flexibility and explore a suitable combination of $\log \rm{g}$ estimates provided by SSPP (we refer the reader to \citealt{Lee08a} for details about different approaches to determine $\log \rm{g}$ from SEGUE spectra). 
We noticed that the distributions of both $\log \rm{g}_{\rm ANNRR}$ and $\log \rm{g}_{\rm SPEC}$ exhibit clearly two peaks, as expected for the BHBs/BSSs dichotomy, while it is not possible to observe this feature in others. These two distributions are shown in the right column of Fig. \ref{fig:gauss}.
Hence, we proceeded with the GMM separation within the two-dimensional space defined by these $\log \rm{g}$ estimates. The final $\log \rm{g}$ adopted by the pipeline was not considered an extra dimension as it consists of a weighted mean of the valid estimates.

In order to guarantee the robustness of our method against uncertainties reported by the SSPP, we constructed a set of $10^4$ realizations of each star's $\log \rm{g}$ estimates in a Monte Carlo framework. Then, we performed the GMM classification for all iterations. Finally, the fraction of instances that a star is attributed to a certain class (either BHB or BSS) is taken as its membership probability for that given group. For this procedure, stars without valid estimates of both $\log \rm{g}_{\rm ANNRR}$ and $\log \rm{g}_{\rm SPEC}$ are removed. With this strategy, we achieved a sample of 5699/4590 stars classified as BHBs above 50\%/99\% probability\footnote{\noindent The full sample is available at \url{https://github.com/guilhermelimberg/bhb_dist}.}.

\begin{figure*}[pt!]
\centering
  \begin{minipage}[b]{0.45\textwidth}
    \centering
    \includegraphics[width=\columnwidth]{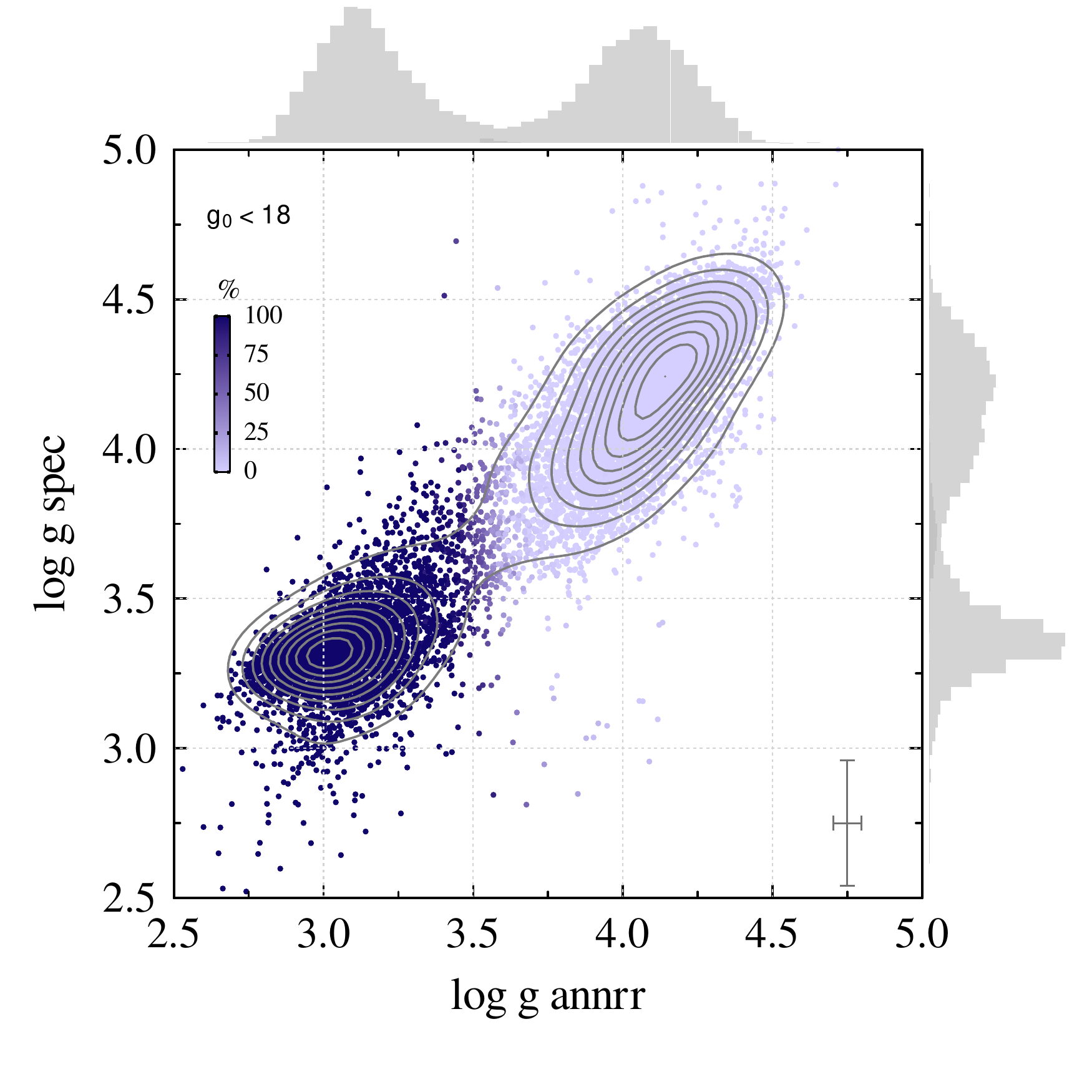}
  \end{minipage}
  \begin{minipage}[b]{0.45\textwidth}
    \centering
    \includegraphics[width=\columnwidth]{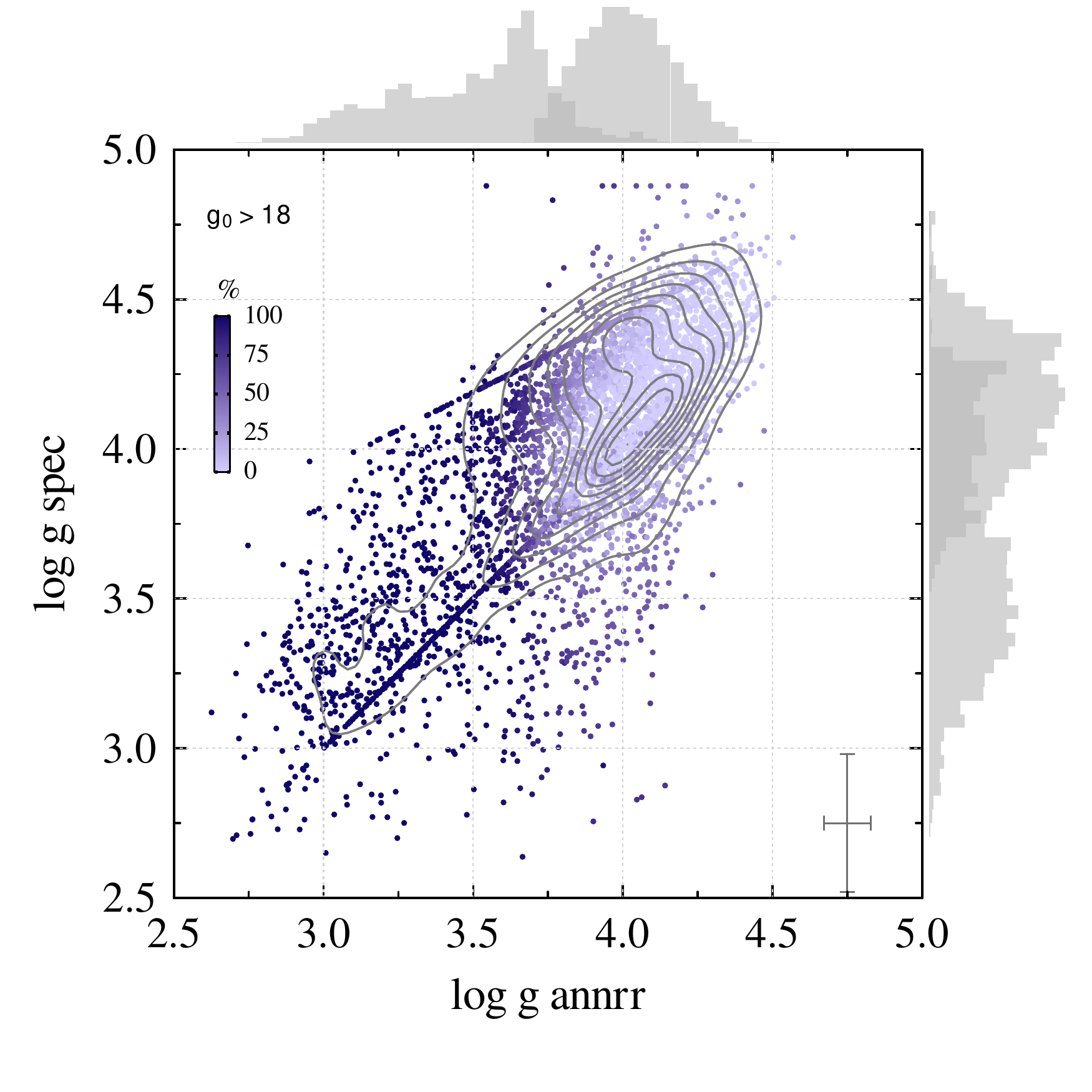}
  \end{minipage}%
  \caption{Distribution of classified stars with $g_0 < 18$ (left) and $g_0 > 18$ (right) in the surface gravity space. Median errors are indicated in the bottom right corner. Histograms show the distribution of log g from the respective axis for stars classified as BHB and BSS. Colors indicate the probability of being a BHB star.}
  \label{fig:class}
\end{figure*}

The final classification obtained is shown in Fig. \ref{fig:class}.
The difference in the uncertainties of the estimates greatly influences the classification, as the $\log \rm{g}_{\rm ANNRR}$ presents more precise values ($\sim 0.06$) than $\log \rm{g}_{\rm SPEC}$ ($\sim 0.21$). 
We cross-matched our sample with the one from \cite{santucci_frac} to evaluate the fraction of BSS contamination. 10\% of our BHB set was classified previously as BSSs, and, among those with a probability greater than 99\% of being BHB following the method implemented here, 2\% is possibly incorrectly assigned.

%===============================================
%===============================================
\section{Absolute magnitude calibration}
\label{sec:calib}

\subsection{Globular clusters stars}
\label{subsec:data_fit}

The procedure to construct an absolute magnitude relation follows previous works (see Section \ref{sec:intro}), starting with the selection of BHB stars in globular clusters. We used the photometric catalog from \cite{an}, which provides magnitudes for crowded fields observed by SDSS. The clusters presenting a well defined horizontal branch were selected and their magnitudes were corrected using the standard extinction ($E(B-V)$) from \cite{sch_ebv} along with the relative extinctions from \cite{wang} for $g-, u-$ and $r-$band.
\cite{vasi} attributed a membership probability for stars in globular clusters based on proper motions and parallaxes from Gaia EDR3. We selected stars from several globular clusters that were more likely than 0.99 to belong to those clusters and we obtained their absolute magnitude in the SDSS $g$-band ($M_g$) with the estimated distance for each cluster given by these authors. 
The list of clusters, their heliocentric distances and distance moduli are presented in Table \ref{tab:cluster}. 

\begin{table}[pb!]
\centering
\caption{Heliocentric distances provided by \citet{vasi} for each globular cluster, uncertainties, and their respective distance moduli.} \label{tab:cluster}
\begin{tabular}{lccc}
\hline
  \multicolumn{1}{c}{Cluster} &
  \multicolumn{1}{c}{D (kpc)} &
  \multicolumn{1}{c}{$\sigma_{D}$ (kpc)} &
  \multicolumn{1}{c}{$(m-M)_0$ (mag)} \\
\hline \hline
  NGC2419 & 83.0 & 1.5 & 19.59 \\
  NGC4147 & 18.65 & 0.16 & 16.35 \\
  NGC5024, M53 & 18.59 & 0.15 & 16.35 \\
  NGC5053 & 17.30 & 0.14 & 16.19 \\
  NGC5272, M3 & 10.20 & 0.06 & 15.04 \\
  NGC5466 & 16.32 & 0.13 & 16.06 \\
  NGC5904, M5 & 7.49 & 0.05 & 14.37 \\
  NGC6205, M13 & 7.53 & 0.06 & 14.38 \\
  NGC6341, M92 & 8.60 & 0.05 & 14.67 \\
  NGC7078, M15 & 10.73 & 0.14 & 15.15 \\
  NGC7089, M2 & 11.62 & 0.13 & 15.33 \\
\hline \hline 
\end{tabular}
\end{table}

To create the sample used to implement the calibration, we applied the limits for colors as defined for the initial selection (see Section \ref{A_stars}). Then, the stars were selected in a single combined CMD, limiting the $M_g$ between $-0.15$ and $1.15$. 
After this exercise, the remaining globular cluster members were checked individually at the SIMBAD database \citep{SIMBAD}, and those classified as variables, blue stragglers and other undesirable types were removed. We also excluded stars with flags in the magnitudes used, leaving us with 744 stars to derive the calibrations from.

\subsection{Fitting the horizontal branch}
\label{subsec:tb}

Finding the best mathematical relationships to fit observable data is not an easy task. In previous works, the absolute magnitudes for BHB stars have been described as a high-degree polynomial \citep{preston,deason,belokurov_bhb}. Instead of arbitrarily assuming that this function is the best representation of the data, we explore the possible combinations between colors and absolute magnitudes. For this task, we employed the \texttt{TuringBot} software \citep{Ashok2020}, a code that performs symbolic regression using a simulated annealing algorithm \citep{delahaye2019,chira2019} in order to search for the best set of parameters and mathematical operations to describe the data.

\texttt{TuringBot} is particularly interesting in this case, because it allows the visualization of the estimated mathematical laws, allowing the user to choose the most appropriate equations for their needs.
Furthermore, the user is free to choose the mathematical operations involved in the fitted functions, the error metric for convergence, as well as the input variables. The best fits are presented in a summarized box, combining the error and the complexity of the equations. The complexity is defined by the sum of the ``size'' of the mathematical operations, constants, and variables present in the solutions. The program assumes that an input variable, constant, sum, subtraction, and multiplication have size 1 each, division has size 2, and more complex operations have higher sizes\footnote{\noindent More \texttt{TuringBot} details can be found in the program documentation, available at: \url{https://turingbotsoftware.com/documentation.html}.}.
We verified that the use of very complex mathematical operations is unnecessary and does not improve the average error of the equations presented by the software. Hence, we adopted only the basic mathematical operations (sum, subtraction, multiplication, and division) as input for the search for absolute magnitude calibrations. The mean absolute error was used as a criterion for convergence and we tested the dependence of all the most common available observable variables found in the literature for estimates of this type, such as magnitudes, color indices, and metallicity.

After evaluating all the combinations of input variables ($u_0$, $g_0$, $r_0$, $(u - g)_0$, $(u -r)_0$, $(g - r)_0$, and [Fe/H]) presented in Appendix \ref{appe}, we found that there is no significant dependence on metallicity in the calibrations provided by the code, regardless of the mathematical operations adopted and the algorithm convergence time. 
The colors $(u - g)_0$ and $(u - r)_0$ provided calibrations with smaller errors than the color $(g - r)_0$, traditionally used in the absolute magnitude calibration of BHB stars (\citetalias{deason}; \citealt{belokurov_bhb}), and also smaller than $(g - r)_0$ with [Fe/H], which means we can achieve more accurate results that do not require metallicity information.
The observed improvement with $(u - g)_0$ and $(u - r)_0$ color might be associated with the $u$ filter, whose transmission curve is mostly between 3000\AA{} and 4000\AA{}, i.e., it is positioned in a region of the spectrum where the Balmer discontinuity ($\sim$3645\AA) is located, as well as several Hydrogen lines from the Balmer series, which makes it a useful indirect indicator of $T_{\rm eff}$ and $\log \rm{g}$ of the BHBs, atmospheric parameters that are directly linked to the mass of the stars in the horizontal branch \citep{valcarce2008}.

\begin{figure}[pt!]
    \centering
    \includegraphics[width=\columnwidth, trim={0 0.5cm 0 0},clip]{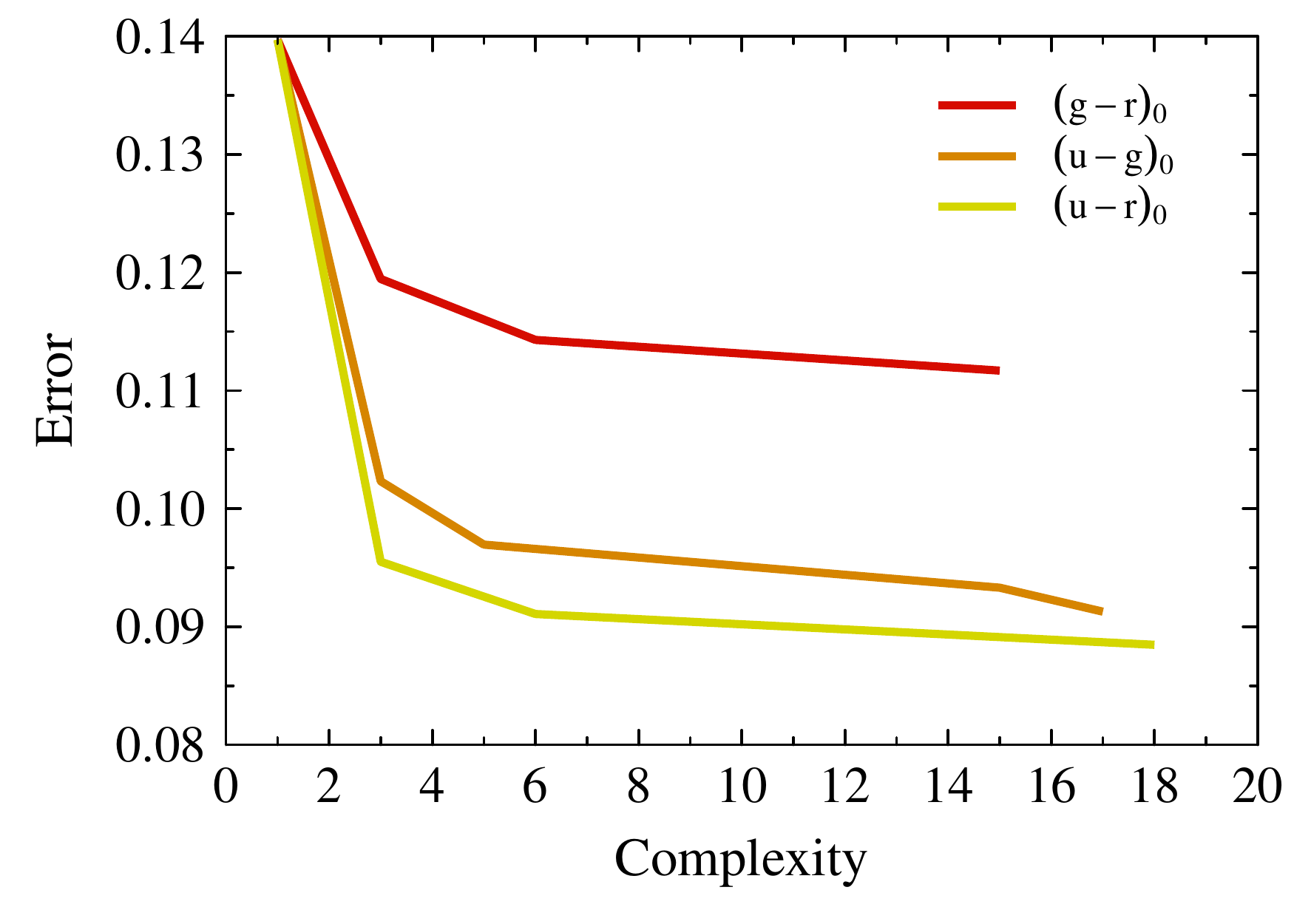}
    \caption{Error comparison between fits for colors $(g - r)_0$, $(u - g)_0$ and $(u - r)_0$ from \texttt{TuringBot}.}
    \label{fig:comp_fits_new}
\end{figure}

Fig. \ref{fig:comp_fits_new} shows the associated errors for each fit in the final BHB sample. Clearly, $(u - r)_0$ presents a better performance than $(g - r)_0$ and $(u - g)_0$. Using the same tool, we find that the relation proposed in \citetalias{deason} is a function of complexity 33 and ${\rm error} = 0.12$, whilst equations of lower order present a much lower complexity with errors of ${\sim}0.10$.

We chose the first functions from which there is no significant decrease in error, i.e., functions of complexity 6 in Fig. \ref{fig:comp_fits_new}, as those that best describes the data.
Exists a singularity in the calibrations, however it is outside of our color range. Hence, it does not imply an obstacle to their usage in the context of this work.

\begin{equation}
\label{eq:gr}
    M_g = \frac{0.178}{0.537 + (g - r)_0}
\end{equation}

\begin{equation}
\label{eq:ur}
    M_g = \frac{0.721}{(u - r)_0} - 0.212
\end{equation}

%===============================================
%===============================================

\subsection{Distances analysis} \label{subsec:dist}

\begin{figure*}[pt!]
  \begin{minipage}[b]{0.5\textwidth}
    \centering
    \includegraphics[width=\columnwidth]{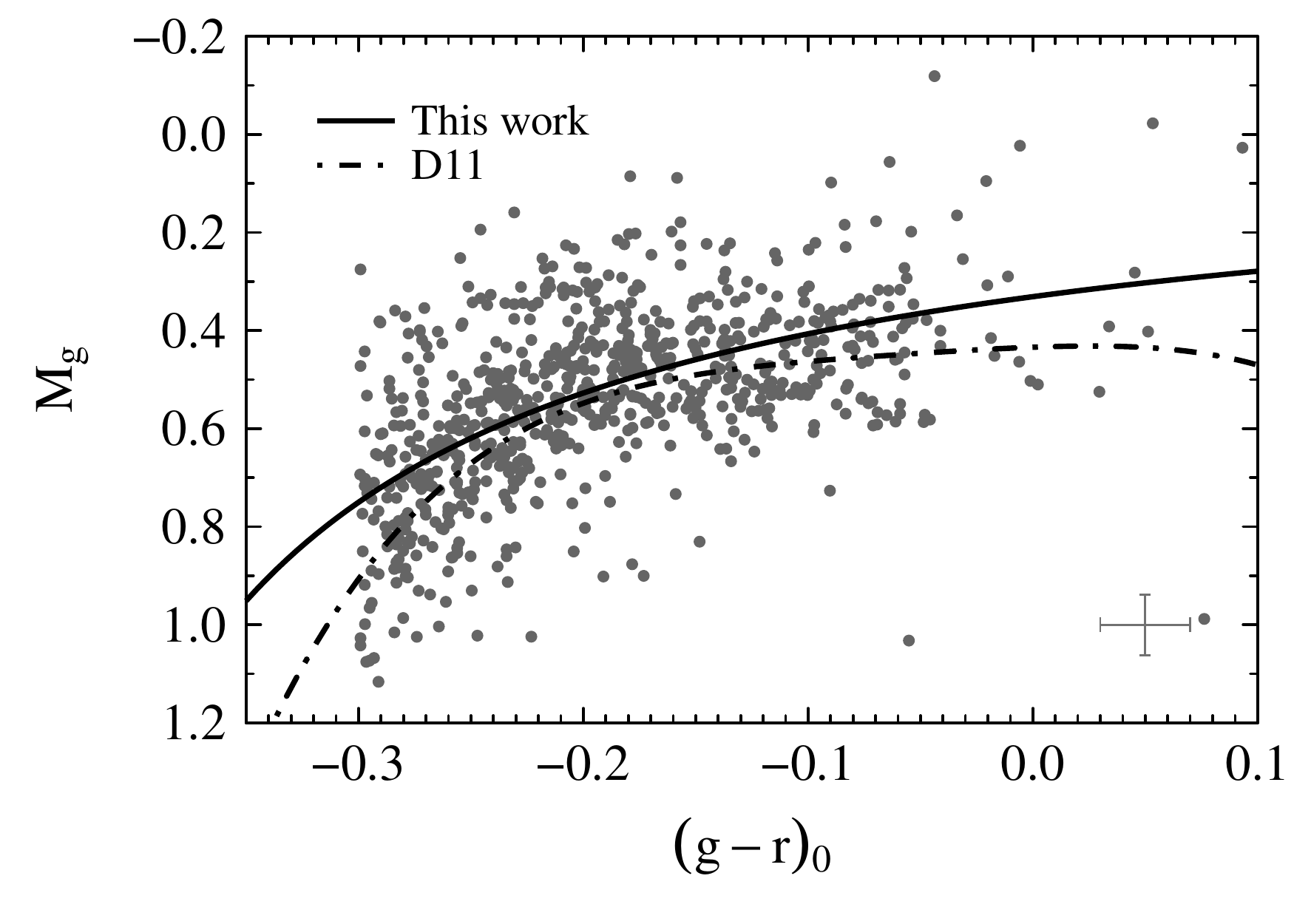}
  \end{minipage}%
  \begin{minipage}[b]{0.5\textwidth}
    \centering
    \includegraphics[width=\columnwidth]{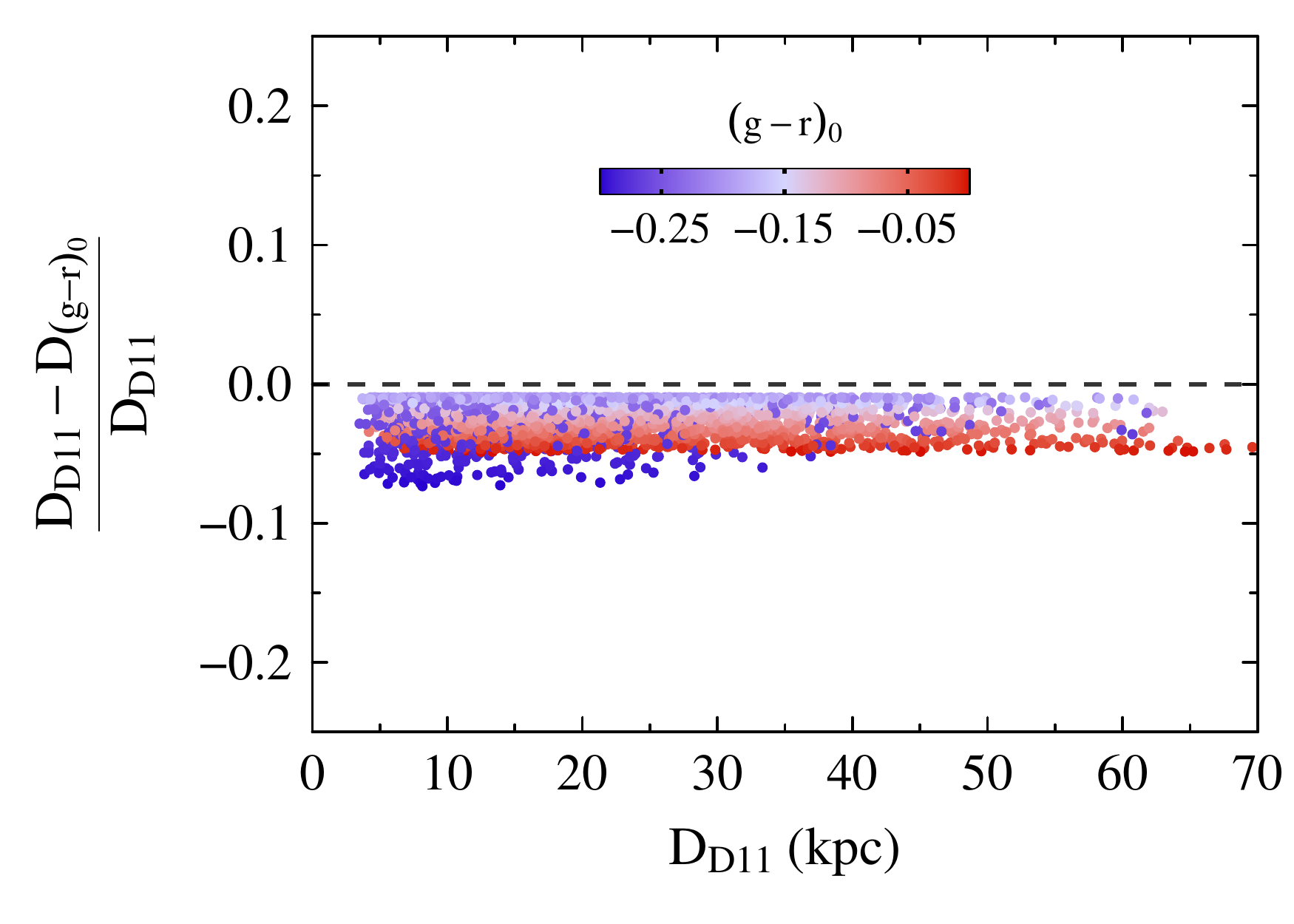}
  \end{minipage}
  \begin{minipage}[b]{0.5\textwidth}
    \centering
    \includegraphics[width=\columnwidth]{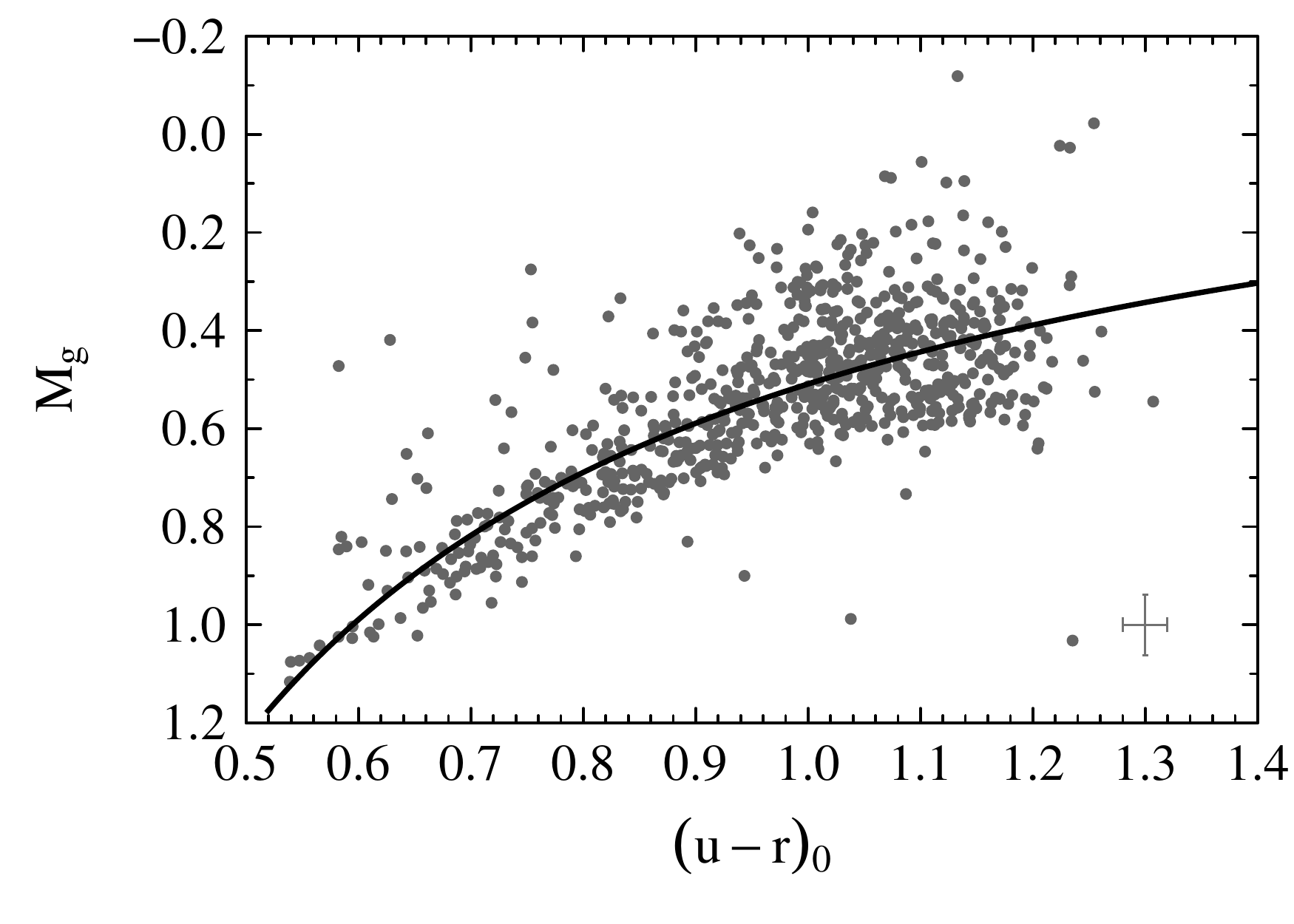}
  \end{minipage}
  \begin{minipage}[b]{0.5\textwidth}
    \centering
    \includegraphics[width=\columnwidth]{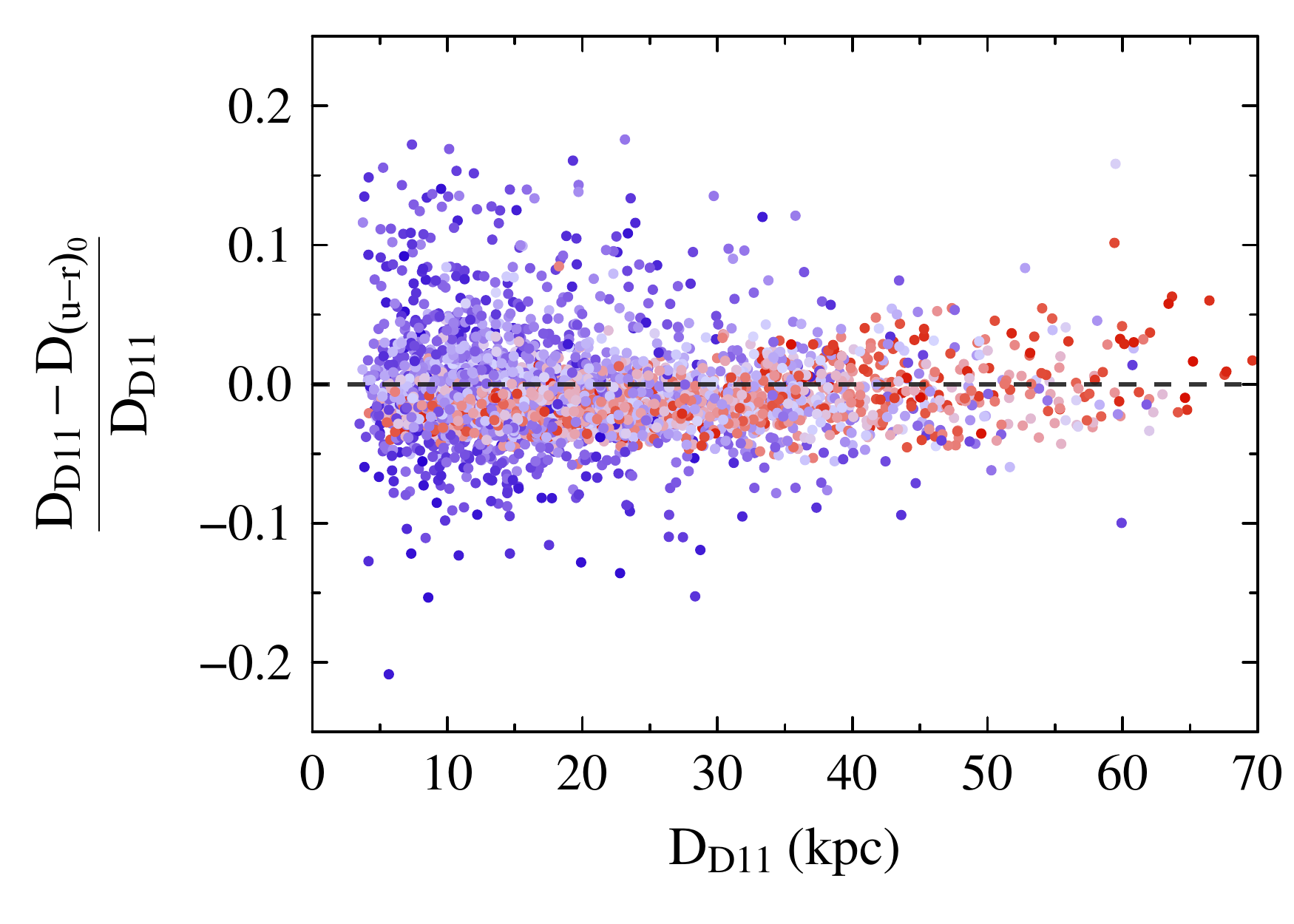}
  \end{minipage}%
  
  \caption{Left panels: color--magnitude diagrams showing BHBs used to define the calibrations. Dash-dotted line represents the polynomial fit defined in \citetalias{deason}. Solid lines represent the calibration for $(g - r)_0$ and $(u - r)_0$ color presented in this work (Eq. \ref{eq:gr} and \ref{eq:ur}, respectively, from top to bottom). Median errors of the data are indicated in the bottom right corner of each panel.
  Right panels: difference between distances calculated with \citetalias{deason}'s calibration and those presented in this work, respectively $(g - r)_0$ (top) and $(u - r)_0$ (bottom).}
  \label{fig:comp_deason}
\end{figure*}

Left panels in Fig. \ref{fig:comp_deason} show the distribution of BHB stars in the CMD with color $(g - r)_0$ (top) and $(u - r)_0$ (bottom). In the top left panel, we can observe how the calibration proposed here (Eq. \ref{eq:gr}) provides magnitudes lower than \citetalias{deason}'s, which results in larger distances. 
The difference is minimal  at $(g - r)_0 \sim -0.20$, where both equation come closer, and the smaller values are a consequence of the inclusion of the cluster NGC7078, whose stars are brighter and were not included in \citetalias{deason}. On the other hand, the distribution using $(u - r)_0$ has a lower dispersion (bottom left panel).

In the right panels (Fig. \ref{fig:comp_deason}), we also show the comparison between distances estimated with the relation from \citetalias{deason} and each calibration defined in the present work. For consistence with \citetalias{deason}'s relation, only stars bluer than $(g - r)_0 = 0$ were considered, and we rejected stars with BHB probabilities of less than 0.99 to reduce the number of misclassified stars. 
The new calibration using color $(g - r)_0$ provides distances about 5\% larger than \citetalias{deason}'s for the reddest stars, while the other end of the color window attains a relative difference of up to 9\% (top right panel).
For the color $(u - r)_0$, the scatter is more uniform and much larger for the bluest stars (bottom right panel).

\begin{figure*}[pt!]
\centering
  \begin{minipage}[b]{\columnwidth}
    \centering
    \includegraphics[width=0.99\columnwidth, trim={-1.05cm 2.55cm 0 0},clip]{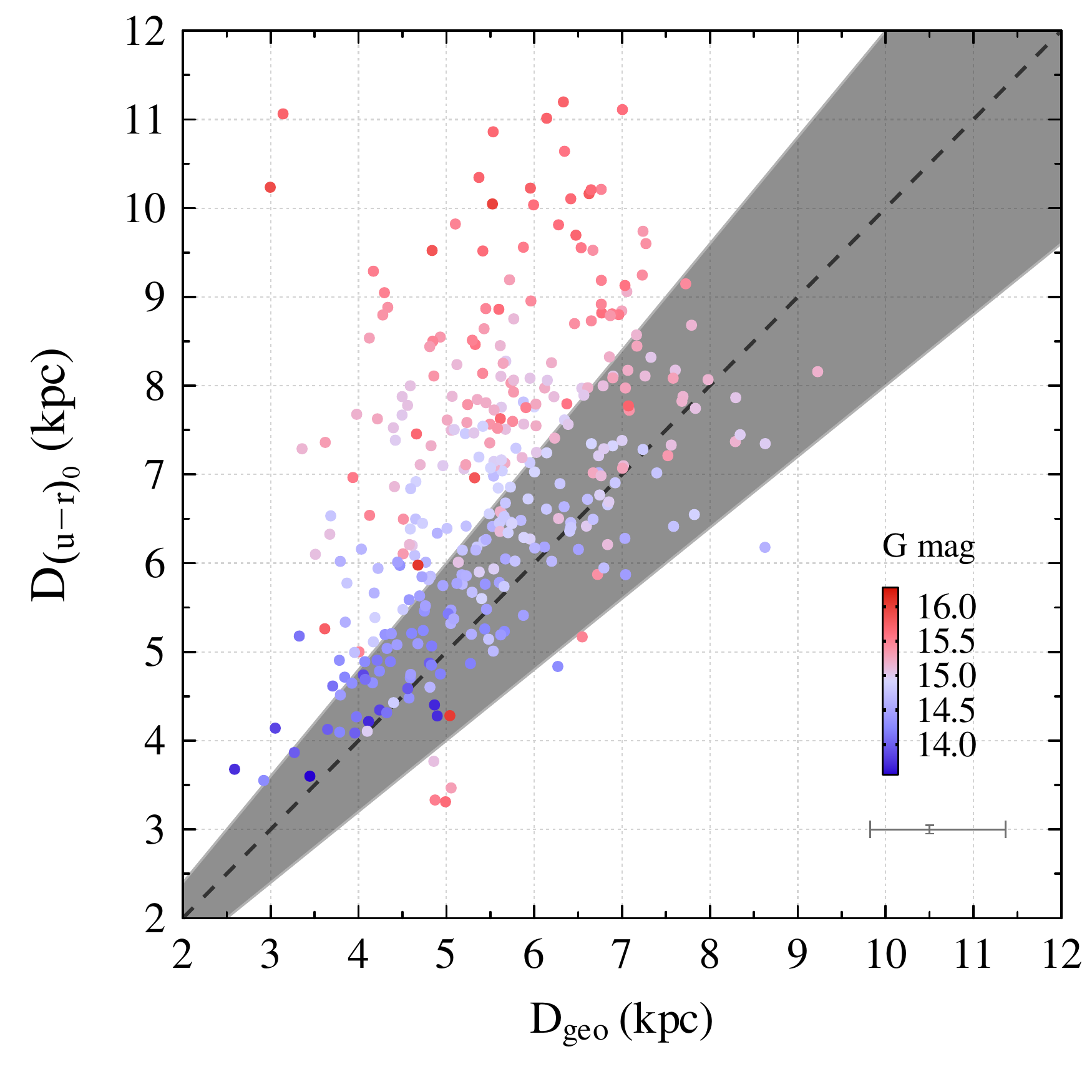}
    \includegraphics[width=\columnwidth, trim={0 0 0 0.2cm},clip]{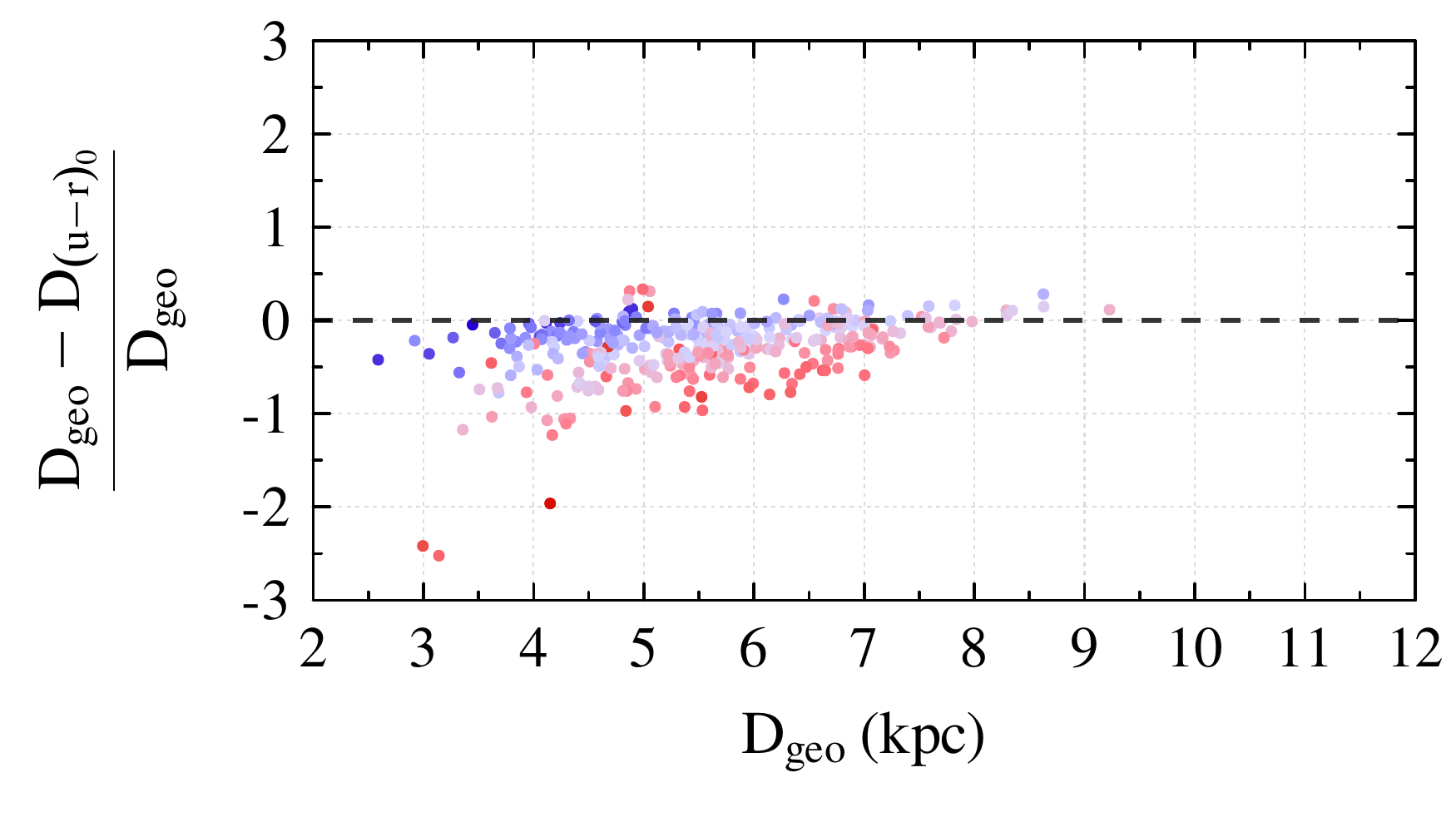}
  \end{minipage}
  \begin{minipage}[b]{\columnwidth}
    \centering
    \includegraphics[width=0.99\columnwidth, trim={-1.05cm 2.55cm 0 0},clip]{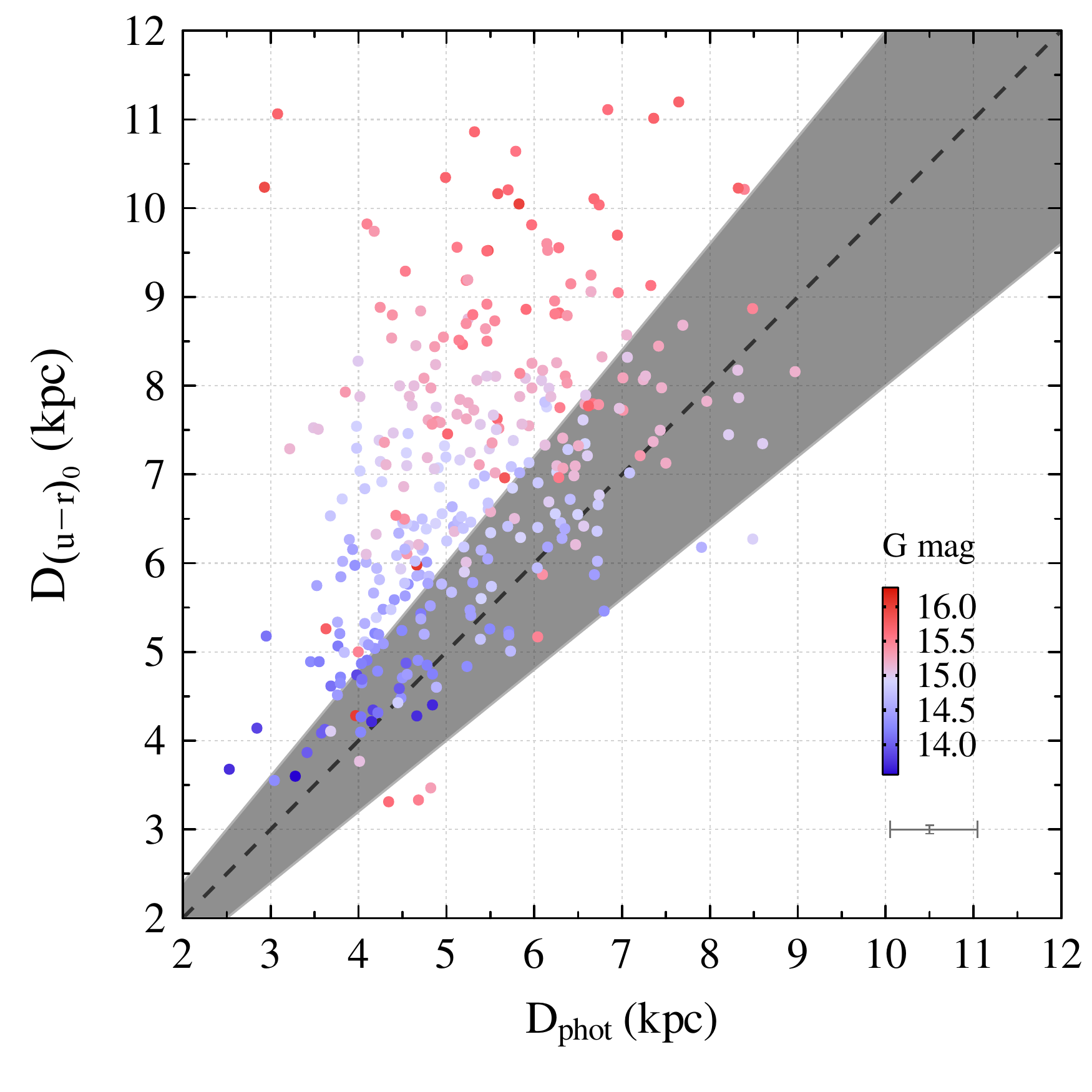}
    \includegraphics[width=\columnwidth, trim={0 0 0 0.2cm},clip]{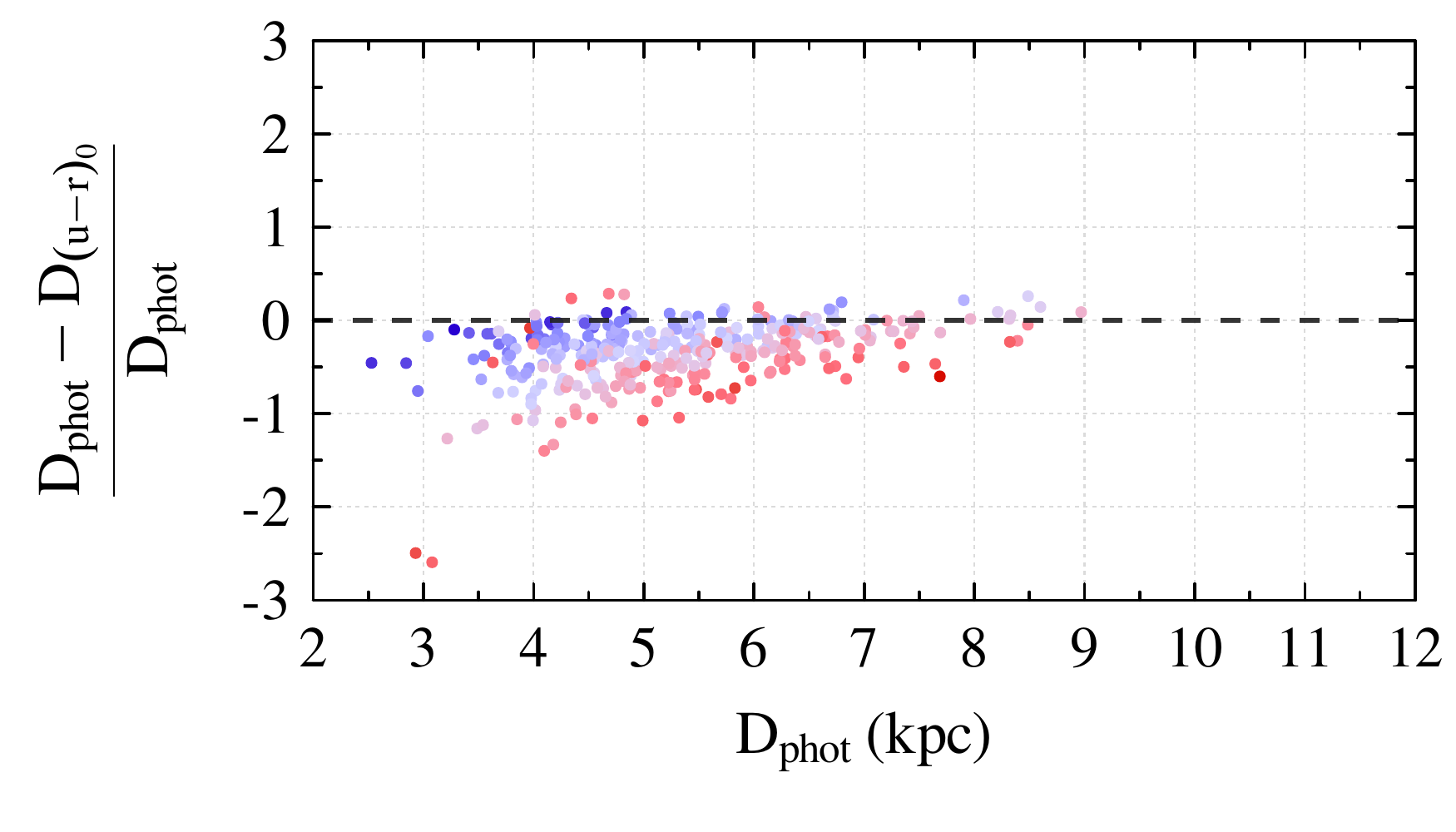}
  \end{minipage}%
  \caption{Comparison between the distances estimated by Eq. \ref{eq:ur} ($D_{(u-r)_0}$), geometric ($D_{geo}$, left) and photogeometric ($D_{phot}$, right) distances from \cite{bailer}. Colors indicate Gaia G magnitude and gray area indicates the region of $1-\sigma$. Median errors are indicated in the bottom right corner of the top panels.}
  \label{fig:comp_bailer}
\end{figure*}

When comparing with purely astrometric heliocentric distances, there is a considerable scatter, even for stars closer than 5 kpc. For this comparison, we selected stars with relative parallax uncertainty from Gaia EDR3\footnote{\cite{gaiadoi}.} in the interval $0 < \sigma_{\varpi}/\varpi < 0.2$, re-normalized unit weight errors within the recommended range ($\texttt{RUWE} < 1.4$; \citealt{ruwe}), and also a BHB probability greater than 0.99 (${>} \ 300$ stars).
In Fig. \ref{fig:comp_bailer}, we show the comparison between geometric (left) and photogeometric (right) distances provided by \cite{bailer} and our calibration using $(u-r)_0$ color. For fainter stars, both \citeauthor{bailer}'s (\citeyear{bailer}) distances are frequently underestimated. 
The gray region indicates the interval within 20\% of distance in the respective horizontal axes, and we find 65\% of the stars when using photogeometric estimates and 54\% with the purely geometric outside it.
Gaia's parallax measurements potentially are not accurate enough for these BHBs and so the final results are not representative of the sample (since the distances inferred from Bayesian methods are strongly dependent on the measured parallax). We also point out that this effect is unlikely the consequence of an inappropriate classification since 94\% of the stars possess $\log \rm{g}_{\rm ADOP} < 3.6$ and would receive the same label by \citeauthor{santucci_frac}'s (\citeyear{santucci_frac}) method.
Finally, we cannot endorse the compatibility between \citetalias{deason}'s distances for BHB stars found in the Pristine survey \citep{pristine} and Gaia DR2's parallaxes. As the Pristine data are not publicly available, it is not possible to evaluate whether the difference is due to the BHB sample used.

Similar inconsistencies between photometric and astrometic-inferred distances were also observed by previous works. Using OB stars, \cite{shull} noted an increase in the discrepancies at ${d > 1.5}$\,kpc, with B-type stars showing smaller values of distances when considering parallaxes alone. Our A-type stars sample seems to follow this same trend. 

%===============================================
%===============================================
\section{Discussion \& summary}
\label{sec:conc}

Since the last absolute magnitude calibration published for BHB stars we had the advent of Gaia data, which allowed us to review the previous relationship thanks to the better characterization of globular clusters \citep{vasi}.

Using data from the SSPP catalog, we obtained a sample of ${\sim}5700$ BHB stars implementing the GMM algorithm. This new approach is an alternative to the previous individual Gaussian fits, as it is clear that they do not distinguish the latest SSPP $\log {\rm g}$ distribution properly (Fig. \ref{fig:gauss}). 

To find which kind of function better describes the distribution of BHBs in the CMD, we used a software for implementing symbolic regression. 
We suggest two new color--magnitude calibrations based on photometry, including a relation with $(u-r)_0$ color that has not been used before. This calibration provides more accurate estimates than $(g-r)_0$ color in most cases. 
For the bluest stars, the differences can exceed 10\% of nominal values (Eq. \ref{eq:ur}). However, this difference decreases for redder and more distant stars.
Here, we show that the calibrations can be simpler and achieve an acceptable result, very similar to those from by \citetalias{deason}'s relation.

We noted substantial differences between photometric and geometric/photogeometric distances. A possibility would be inaccurate estimates of $\log {\rm g}$ provided by SEGUE, as it was mainly designed for cool stars ($T_{\rm eff} < 7500\,{\rm K}$). However, \cite{santucci_frac} showed that this is unlikely to be the case. It could also be due to an incorrect value of extinction for the SDSS photometry, yet it also does not seem to explain the disparity. Most stars were observed in regions of low extinction and we could not find any relation between the extinction values and the inconsistency observed.
The observed differences also lead us to believe that measured parallaxes for these stars could be unreliable, as there is no agreement between the distance estimates even for the closest stars. It would be interesting to investigate whether the same discrepancy can be observed with other halo tracers.

The new sample made available here can help to improve results already known about the structures of the Milky Way stellar halo. For example, these BHBs can be used to revisit the duality of the stellar halo or re-evaluate the Galaxy's mass estimate.

%===============================================
%===============================================

\acknowledgments

We thank the anonymous referee for the careful review and all the suggestions, which greatly improved our work. % definitely the best referee I will ever have
This research was financed with public funds, without which it would not have been possible. 
F.O.B. acknowledges CAPES (PROEX; Proc. 88887.604787/2021-00). R.M.S. acknowledges CNPq (Proc. 306667/2020-7). S.R. acknowledges partial financial support from FAPESP (Proc. 2015/50374-0 and 2014/18100-4), CAPES, and CNPq. G.L. acknowledges FAPESP (Proc. 2021/10429-0). A.P.-V. acknowledges the DGAPA-PAPIIT grant IA103122. H.D.P. thanks FAPESP (Procs. 2018/21250-9 and 2022/04079-0).

% Gaia
This work has made use of data from the European Space Agency (ESA) mission Gaia (\url{https://www.cosmos.esa.int/gaia}), processed by the Gaia Data Processing and Analysis Consortium (DPAC, \url{https://www.cosmos.esa.int/web/gaia/dpac/consortium}). Funding for the DPAC has been provided by national institutions, in particular the institutions participating in the Gaia Multilateral Agreement.

% SDSS
Funding for the Sloan Digital Sky Survey IV has been provided by the Alfred P. Sloan Foundation, the U.S. Department of Energy Office of Science, and the Participating Institutions. SDSS acknowledges support and resources from the Center for High-Performance Computing at the University of Utah. The SDSS web site is www.sdss.org.
SDSS is managed by the Astrophysical Research Consortium for the Participating Institutions of the SDSS Collaboration including the Brazilian Participation Group, the Carnegie Institution for Science, Carnegie Mellon University, Center for Astrophysics | Harvard \& Smithsonian (CfA), the Chilean Participation Group, the French Participation Group, Instituto de Astrofísica de Canarias, The Johns Hopkins University, Kavli Institute for the Physics and Mathematics of the Universe (IPMU) / University of Tokyo, the Korean Participation Group, Lawrence Berkeley National Laboratory, Leibniz Institut für Astrophysik Potsdam (AIP), Max-Planck-Institut für Astronomie (MPIA Heidelberg), Max-Planck-Institut für Astrophysik (MPA Garching), Max-Planck-Institut für Extraterrestrische Physik (MPE), National Astronomical Observatories of China, New Mexico State University, New York University, University of Notre Dame, Observatório Nacional / MCTI, The Ohio State University, Pennsylvania State University, Shanghai Astronomical Observatory, United Kingdom Participation Group, Universidad Nacional Autónoma de México, University of Arizona, University of Colorado Boulder, University of Oxford, University of Portsmouth, University of Utah, University of Virginia, University of Washington, University of Wisconsin, Vanderbilt University, and Yale University.

This research has also made use of RStudio \citep{rstudio} and TOPCAT (\url{http://www.starlink.ac.uk/topcat/}, \citealp{topcat}).

\bibliographystyle{aasjournal}
\bibliography{bibliography.bib}

\appendix
\section{Fitting the horizontal branch with [Fe/H]}
\label{appe}

For the analysis of dependence of metallicity, we have a smaller sample of stars than the one used for the calibrations, about 40 stars from the original sample were available in the SSPP data. The same procedure was done with both the pure photometric sample and this one cross-matched with SSPP.

Each solution evaluation (considering different input variables) was taken in a period of approximately 10\,min, which is enough for the convergence of several solutions, as \texttt{TuringBot} needs less time to converge than other similar softwares \citep{Ashok2020}. The hardware involved in this process is highly important for the convergence time. In our case, the program was executed in a computer with an AMD Ryzen 7 2700 processor, with 16 threads, all used at once.
Tests were also made by running the program longer and no significant improvement was observed.

\begin{figure}[H]
    \centering
    \includegraphics[width=0.8\columnwidth]{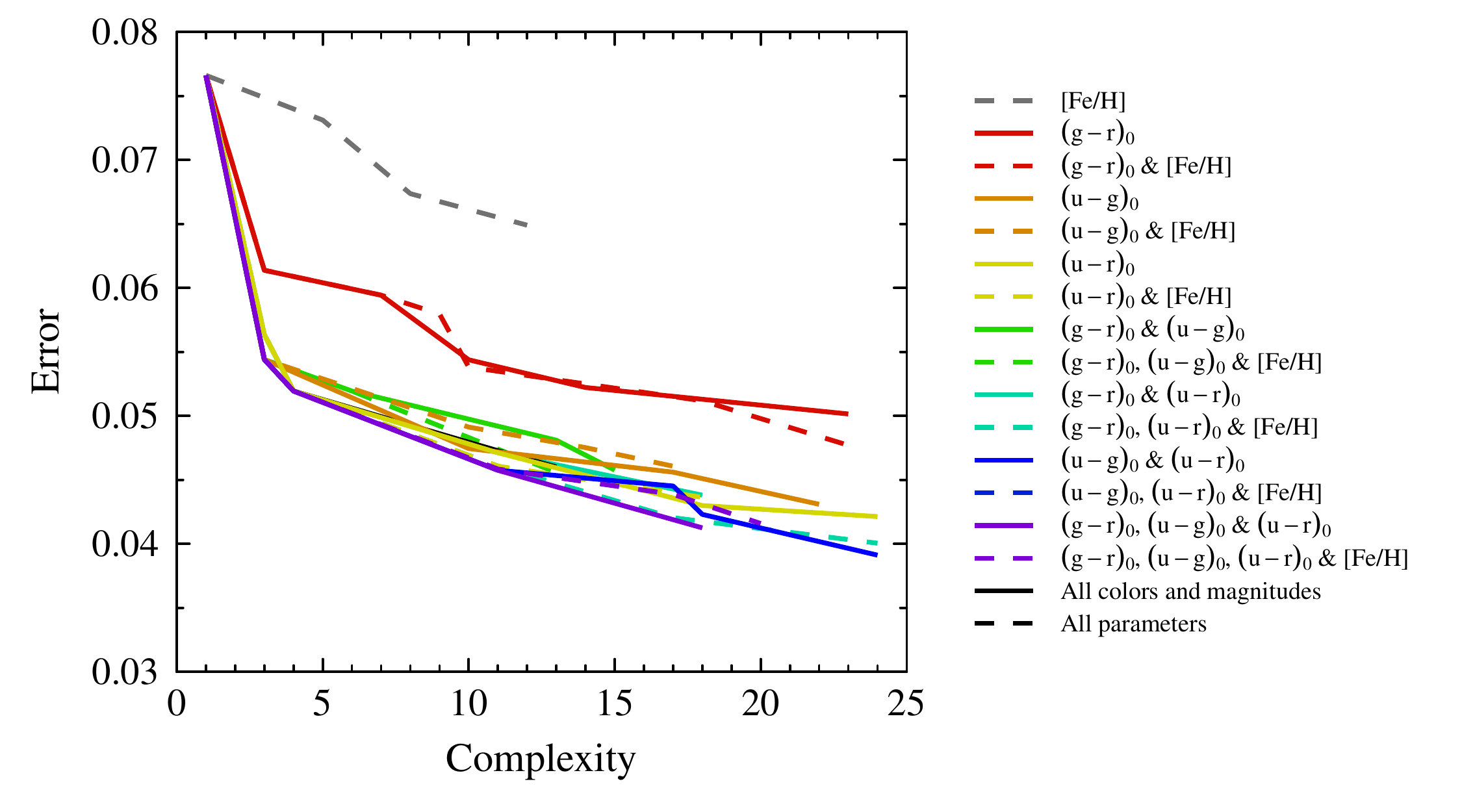}
    \caption{Error comparison between fits for some combinations of magnitudes, colors, and [Fe/H] from \texttt{TuringBot}.}
    \label{fig:comp_fits_fe}
\end{figure}

Figure \ref{fig:comp_fits_fe} shows how the error decreases with the increase of the complexity  of the functions (see Section \ref{subsec:tb} for the definition of ``complexity''). Some fits coincide since the program can create colors from magnitudes, if it is better than the magnitudes alone, and not use all the variables provided. This is the reason why the line for all the parameters is not visible. We can see that, for equations up to complexity 25, there is no advantage of adding the metallicity information as we can achieve similar errors using only magnitudes.

\section{Globular cluster distances}
\label{apB}

\begin{figure}[H]
    \centering
    \includegraphics[width=0.8\columnwidth]{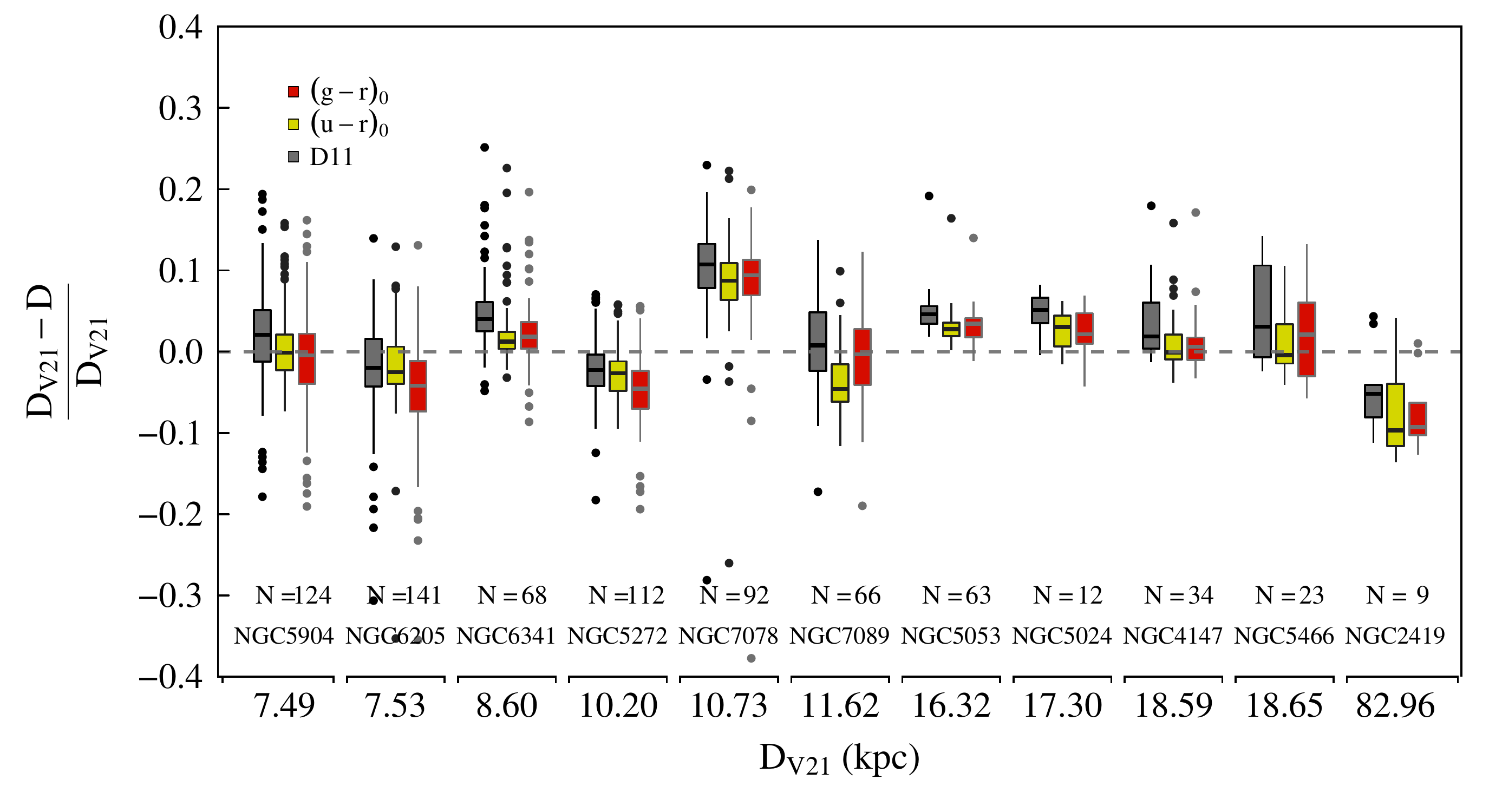}
    \caption{Boxplot for relative distance difference using each calibration presented in this work (Eq. \ref{eq:gr} and Eq. \ref{eq:ur}) and that from \citetalias{deason}. $D_{V21}$ is the distance provided by \cite{vasi}, inferred from Gaia EDR3 data. The number of data points in each cluster is displayed at the bottom of the panel.}
    \label{fig:comp_gc}
\end{figure}

Figure \ref{fig:comp_gc} presents a boxplot for the distances obtained with both calibrations and the one from \citetalias{deason} compared to those provided by \cite{vasi}. The size of each box represent the spread (25th and 75th quartiles) and the center line indicates the median value. The number of stars in each cluster is displayed above their NGC identifier. 

The overall results indicate that Eq. \ref{eq:ur} is in general more accurate than \citetalias{deason}'s relation. This conclusion is supported by better distance predictions where 10 of the 11 globular clusters were better constrained with Eq. \ref{eq:ur}. The calibration with color $(u-r)_0$ revealed to be more accurate than using the relation provided in this work for color $(g-r)_0$, as nine clusters present lower dispersion and eight clusters show medians close to zero when using the former color. NGC2416 was the only cluster where the color $(g-r)_0$ presented a better performance, however, this cluster was represented by only 9 members and therefore this result may be due to sub-sampling. 
The sample considered has a bias to closer cluster, which is apparent by huge distance gap between NGC2416, the farthest cluster, and NGC5466, the penultimate. Therefore, we cannot ascertain whether or not the accuracy of distances estimated using color $(u-r)_0$ varies with the distance.

%% >> ficou MUITO boa essa figura!! (:

\end{document}